\documentclass[a4paper,11pt]{article}
\usepackage{pos}

\usepackage{amsmath}

\newcommand{\bfrac}[2]{\left(\frac{#1}{#2}\right)}
\newcommand{\infrac}[2]{{#1}/{#2}}
\newcommand{\inbfrac}[2]{\left({#1}/{#2}\right)}
\newcommand{\derr}[2]{\frac{d{#1}}{d{#2}}}

\newcommand{\txt}[1]{\textrm{#1}}

\makeatletter
\newsavebox\myboxA
\newsavebox\myboxB
\newlength\mylenA
\newcommand{\barr}[2][1.0]{%
	\sbox{\myboxA}{$\m@th#2$}%
	\setbox\myboxB\null
	\ht\myboxB=\ht\myboxA%
	\dp\myboxB=\dp\myboxA%
	\wd\myboxB=#1\wd\myboxA
	\sbox\myboxB{$\m@th\overline{\copy\myboxB}$}
	\setlength\mylenA{\the\wd\myboxA}
	\addtolength\mylenA{-\the\wd\myboxB}%
	\ifdim\wd\myboxB<\wd\myboxA%
	\rlap{\hskip 0.5\mylenA\usebox\myboxB}{\usebox\myboxA}%
	\else
	\hskip -0.5\mylenA\rlap{\usebox\myboxA}{\hskip 0.5\mylenA\usebox\myboxB}%
	\fi}
\makeatother

\title{Cumulants: It's More Than You Think}
\ShortTitle{Cumulants: It's More Than You Think}

\author*[a]{Agnieszka Sorensen}
\author[b]{Dmytro Oliinychenko}
\author[c]{Volker Koch}
\author[b]{Larry McLerran}

\affiliation[a]{Department of Physics and Astronomy, University of California, Los Angeles, CA 90095, USA}

\affiliation[b]{Institute for Nuclear Theory, University of Washington, Box 351550, Seattle, Washington 98195, USA}

\affiliation[c]{Lawrence Berkeley National Laboratory, 1 Cyclotron Road, Berkeley, California 94720, USA}

\emailAdd{agnieszka.sorensen@gmail.com}

\abstract{Cumulants of baryon number are given considerable attention in analyses of heavy-ion collision experiments as possible signatures of the QCD critical point. In this work, we show that the values of the lowest three cumulants can also be utilized to recover information about the isothermal speed of sound and its logarithmic derivative with respect to the baryon number density. This result provides a new method for obtaining information about fundamental properties of nuclear matter studied in heavy-ion collisions, with consequences for both the search for the QCD critical point and neutron star studies. While the approximations and the model comparison we considered apply to experiments at low energies, the approach itself can be used at any collision energy provided that measurements of cumulants of baryon number distribution as well as their temperature dependence are available.}

\FullConference{%
	The International conference on Critical Point and Onset of Deconfinement - CPOD2021\\
  	15 – 19 March 2021\\
 	Online - zoom
 }

\begin{document}
\maketitle

\section{Introduction}

The speed of sound $c_s$ is one of the fundamental macroscopic properties of any substance. In the case of dense nuclear matter, the behavior of $c_s$ is of direct interest to both the heavy-ion collision and the neutron star research communities. In studies devoted to the QCD phase diagram at high temperature $T$, probed in accelerator experiments colliding heavy-ions, a minimum of the speed of sound signals a phase transition from hadron to quark and gluon degrees of freedom; in particular, the speed of sound is expected to become exactly zero at the critical point (CP) and on the associated spinodal lines of the first-order phase transition. First-principle studies using lattice QCD show that $c_s$ in matter at zero baryon chemical potential $\mu_B$ exhibits a minimum around a pseudocritical temperature of $T_0=156.5\pm1.5$ MeV \cite{Bazavov:2018mes} (see also \cite{Borsanyi:2020fev}), which implies a crossover transition between the hadron gas and the quark-gluon plasma (QGP); whether this transition becomes a first-order phase transition at finite values of $\mu_B$ is currently a subject of intense experimental and theoretical effort.  In neutron star studies, dealing mostly with dense nuclear matter at zero temperature, the behavior of the speed of sound as a function of the baryon density $n_B$ is directly connected to the equation of state (EOS) of nuclear matter at $T=0$, which in turn determines the mass-radius relationship of neutron stars \cite{Ozel:2016oaf}. 

Observations of neutron stars with masses around and above $2M_{\odot}$ \cite{Demorest:2010bx,Antoniadis:2013pzd,NANOGrav:2019jur}, where $M_{\odot}$ is the mass of the Sun, suggest that the EOS of dense nuclear matter is considerably stiff at densities larger than 2-3 times the saturation density of nuclear matter $n_0$, which in turn has led a number of authors to suggest that $c_s$ in dense nuclear matter may even exceed the conformal limit of $1/\sqrt{3}$ \cite{Bedaque:2014sqa,Tews:2018kmu,McLerran:2018hbz}. This possibility has been further strengthened by a recent machine learning study \cite{Fujimoto:2019hxv}, where the known population of neutron stars is used to infer $c_s$ as a function of energy density, and which suggests that the value of $c_s$ may exceed $1/3$ substantially.

To our best knowledge, only a few attempts have been made to estimate $c_s$ in matter created in heavy-ion collisions \cite{Gardim:2019xjs, Steinheimer:2012bp}. In our study \cite{Sorensen:2020ygf}, we suggest a novel way of measuring $c_s$ using well-known heavy-ion collision observables: the cumulants of the baryon number distribution, which are central observables in studies devoted to the QCD phase diagram due to their characteristic behavior near the CP \cite{Asakawa:2009aj, Stephanov:2011pb}. This qualitatively distinct behavior of the cumulants near the CP follows directly from their dependence on the derivatives of pressure $P$ with respect to $\mu_B$. The key insight used in our study is that the cumulants are sensitive to the behavior of the EOS everywhere in the phase diagram, and in particular they can convey information about the magnitude and baryon-density--dependence of the speed of sound.

\section{The relationship between the cumulants and the speed of sound}

Cumulants of the net baryon number $\kappa_j$ are defined as $\kappa_j=VT^{j-1}\big(d^jP/d\mu_B^j\big)_T$, where $V$ is the volume. Expressed in terms of derivatives with respect to $n_B$, the first three cumulants are
\begin{eqnarray}
&& \kappa_1  = V n_B ~, \label{cumulant_1} \hspace{8mm} \kappa_2 = \frac{VTn_B}{\left( \frac{dP}{dn_B} \right)_T}~, \hspace{8mm} \kappa_3= \frac{VT^2n_B}{\left( \frac{dP}{dn_B} \right)_T^2} \left[  1 - \frac{n_B}{\left( \frac{dP}{dn_B} \right)_T} \left(\frac{d^2P}{dn_B^2}\right)_T \right] \label{cumulant_3} ~.
\end{eqnarray}
At the same time, cumulants can be directly computed from the moments of the net baryon number, available in experiment; in particular, for $j\leq3$ we have $\kappa_j\equiv\big\langle\big(N_B-\big\langle N_B\big\rangle\big)^j\big\rangle$.

The square of the speed of sound is equal to the change in pressure induced by a change in energy density, $c_s^2 \equiv \infrac{dP}{d\mathcal{E}}$. Importantly, calculating the speed of sound requires specifying which thermodynamic variables are allowed to vary and which are kept constant. For example, the propagation of sound in air is an adiabatic process, so that the speed of sound in air has to be calculated at constant entropy; in particular, in relativistic problems it is necessary to include the possible creation and annihilation of particles, and as a result the entropy per baryon, $\sigma = S /N_B$, has to be kept constant, leading to the adiabatic speed of sound given by $c_{\sigma}^2 = \inbfrac{dP}{d\mathcal{E}}_\sigma $. On the other hand, if the considered problem involves a temperature reservoir (which is the case, e.g., in porous media) or other means of maintaining constant temperature (e.g., through radiative cooling), one calculates the isothermal speed of sound, $c_T^2 \equiv \inbfrac{dP}{d\mathcal{E}}_T$. 

In this work we will focus in $c_T^2$, which can be written explicitly in terms of derivatives of thermodynamic observables with respect to $n_B$ and then further rewritten using the cumulants,
\begin{eqnarray}
c_T^2 = \bfrac{dP}{d\mathcal{E}}_T = \frac{\Big( \frac{dP}{dn_B} \Big)_T}{ T  \Big(\frac{d s}{d n_B} \Big)_T   +  \mu_B  } = \left[ \frac{T}{\kappa_1}\left(\derr{\kappa_1}{T}\right)_{\mu_B} + \frac{\mu_B}{T} \frac{\kappa_2}{\kappa_1} \right]^{-1}~.
\label{speed_isothermal}
\end{eqnarray}
The first term of the final expression in the above equation can be neglected for $\inbfrac{T}{\mu_B} \ll 1$, leaving an approximate expression for $c_T^2$ in terms of observables available from the experiment,
\begin{eqnarray}
c_T^2 \approx \frac{T \kappa_1}{\mu_B \kappa_2}~.
\label{magic_equation_1}
\end{eqnarray}
We note that Eq.\ (\ref{magic_equation_1}) provides an upper limit to the value of $c_T^2$. Additionally, as is clear from Eq.\ \eqref{speed_isothermal}, in the limit $T \to 0$ we have $c_T^2\big|_{T=0} = \inbfrac{1}{\mu_B} \big( \infrac{dP}{dn_B} \big)_T $; noteworthy, the $T \to 0$ limit for $c_\sigma^2$ is the same, and consequently, for $\inbfrac{\mu_B}{T} \ll 1$, we have $c_{\sigma}^2 \approx c_T^2$.

Eq.\ \eqref{speed_isothermal} also allows us to compute the logarithmic derivative of the speed of sound, which can likewise be expressed in terms of the cumulants as
\begin{eqnarray}
\bigg(\frac{d \ln c_T^2}{d \ln n_B} \bigg)_T + c_T^2
= 1 - \frac{\kappa_3 \kappa_1}{\kappa_2^2}   - c_T^2 \bigg(\frac{d \ln (\kappa_2/T)}{d \ln T}\bigg)_{n_B} ~.
\end{eqnarray}
For $\inbfrac{\mu_B}{T} \ll 1$, we can neglect the last term on the right-hand side, leading to
\begin{eqnarray}
\left(\frac{d \ln c_T^2}{d \ln n_B} \right)_T + c_T^2  \approx 1 - \frac{\kappa_3 \kappa_1}{\kappa_2^2} ~.
\label{magic_equation_2}
\end{eqnarray}
For both Eq.\ \eqref{magic_equation_1} and Eq.\ \eqref{magic_equation_2}, the correction due to the discarded terms is of order $\left(T/\mu_B\right)^2$.

Finally, we note that in the opposite limit, for $\mu_B \to 0$, Eq.\ \eqref{speed_isothermal} becomes \begin{eqnarray}
c_T^{2} = \left( \frac{T}{\kappa_2} \frac{d \kappa_2}{d  T} \right)^{-1}_{\mu_B=0} ~. 
\label{magic_equation_3}
\end{eqnarray}
As a result, provided measurements of $\kappa_2$ are available at different temperatures (which could possibly be achieved by using data obtained at different centralities, energies, and rapidity ranges), one could attempt estimating $c_T^2$ in ultrarelativistic heavy-ion collisions.

At a first glance, the formulas in Eqs.\ \eqref{magic_equation_1}, \eqref{magic_equation_2}, and \eqref{magic_equation_3} may seem counter-intuitive as the speed of sound of the system seems to be driven entirely by baryon number fluctuations; this may be perceived to be in contrast to, e.g., results from lattice QCD, applicable for $\mu_B \approx 0$, where the value of $c_{\sigma}^2$ is dominated by contributions from pions. To address this apparent contradiction, let us consider a simplified scenario of a gas of protons, antiprotons, and pions for a specific case of $\mu_B \approx 0$. In this case using $\mu_B$ as the auxiliary variable in the chain rule is more intuitive, and it can be easily seen that starting from $c_T^2 = \inbfrac{dP}{d\mu_B}_T / \inbfrac{d\mathcal{E}}{d\mu_B}_T$ leads to the same result as in Eq.\ \eqref{speed_isothermal}, so that in the limit $\mu_B \to 0$ we are again led to  Eq.\ \eqref{magic_equation_3}. We then note that when $T$ is kept constant, any fluctuations in the system, including those involved in the propagation of sound, must originate from fluctuations in $\mu_B$; in other words, the density of pions, governed by the value of $T$, remains fixed, but the net proton density is allowed to oscillate around it's value $n_P(\mu_B \approx0) \approx 0$. As a result, $c_T^2$ is driven solely by the value of $T$ and the fluctuations in the net proton number, in agreement with Eq.\ \eqref{magic_equation_3}. This is in contrast to the case of $c_{\sigma}^2$, which at $\mu_B \approx 0$ is given by a simple expression $c_{\sigma}^2 = s/ \left( T ds/dT\right) = s/c_V$, where $c_V$ is the specific heat at constant volume. The entropy of a gas of protons, antiprotons, and pions at $\mu_B \approx 0$ is naturally dominated by pions, and consequently the contribution due to pions dominates the value of $c_{\sigma}^2$.

\section{Validation}

Below, we focus on the application of formulas obtained in Eqs.\ \eqref{magic_equation_1} and \eqref{magic_equation_2}, and we first test the range of $T$ and $\mu_B$ in which they are good approximations of exact expressions for $c_T^2$ and its logarithmic derivative. To this end, we use two models of dense nuclear matter. First of them is the well-known Walecka model \cite{Walecka:1974qa}, which utilizes interactions of scalar and vector type and describes the ordinary nuclear liquid-gas phase transition. The other model is the relativistic vector density functional (VDF) model with two phase transitions \cite{Sorensen:2020ygf}, which utilizes vector-type interaction terms and describes two phase transitions: the nuclear liquid-gas phase transition and a conjectured high-density, high-temperature phase transition modeling the QGP phase transition. The position of the critical point and the width of the spinodal region associated with this ``QGP-like'' phase transition can be freely chosen within the VDF model; in this work, the QGP-like phase transition is chosen to exhibit a CP at $T_c=100\ \txt{MeV}$ and  $n_c=3n_0$, with the $T=0$ boundaries of the spinodal region in the $T$-$n_B$ plane given by $n_{B, \txt{left spinodal} } (T=0) \equiv \eta_L=2.5 n_0$ and $n_{B, \txt{right spinodal}} (T=0) \equiv\eta_R=3.32 n_0$, where $n_0=0.160\ \txt{fm}^{-3}$; this choice is arbitrary and serves as a plausible example.

\begin{figure}[t]
	\includegraphics[width = 0.99\textwidth]{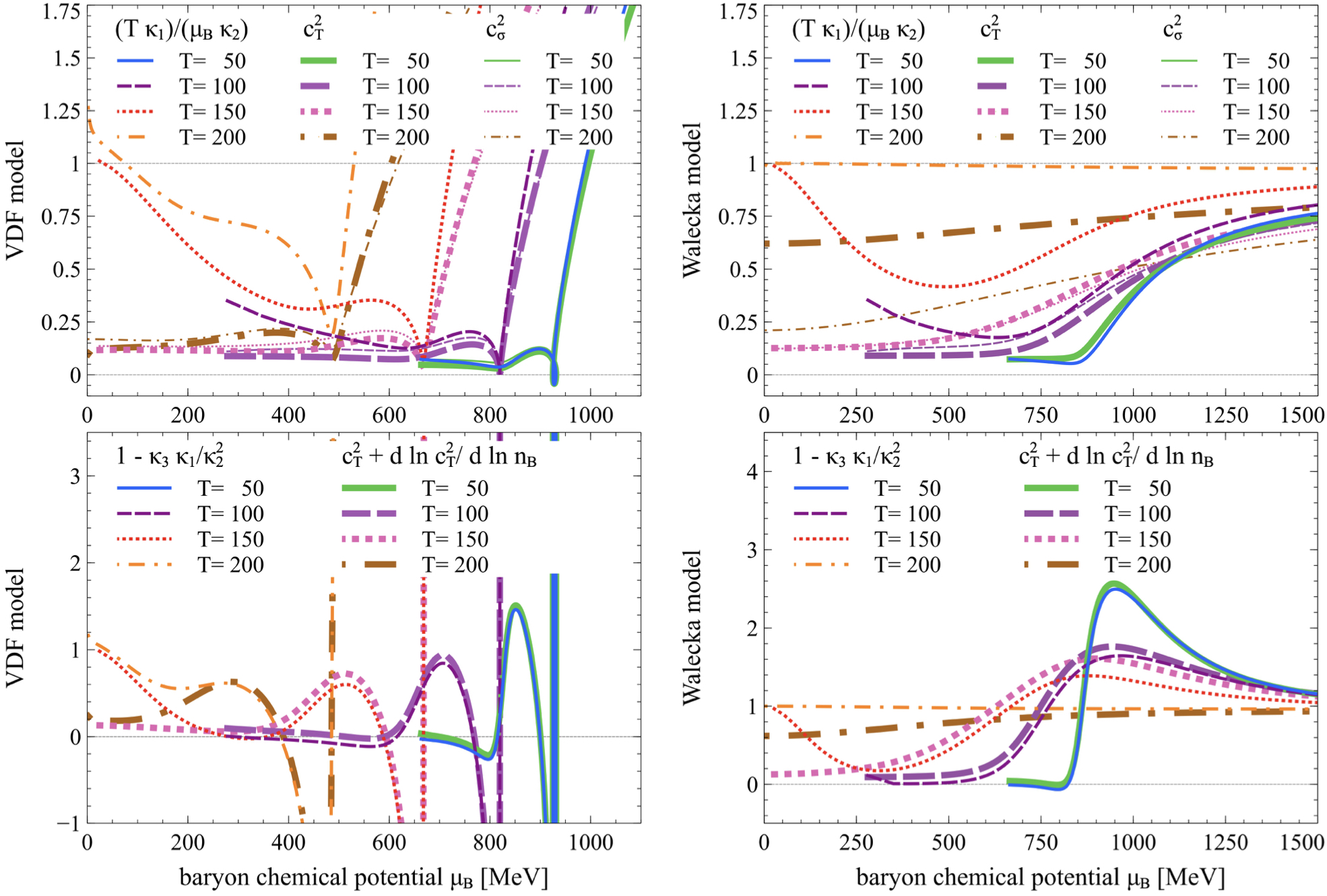}
	\caption{Model study of regions of applicability of Eqs.\ (\ref{magic_equation_1}) and (\ref{magic_equation_2}). The left (right) panels show results obtained in the VDF (Walecka) model. The upper and lower panels show quantities entering Eq.\ (\ref{magic_equation_1}) and Eq.\ (\ref{magic_equation_2}), respectively. Results at $T=50,100,150,200\ \txt{MeV}$ are given by blue and green solid lines, dark and light purple long-dashed lines, red and pink short-dashed lines, and orange and brown dash-dotted lines, respectively. For each $T$, the thickest lines correspond to the exact results and the medium-thick lines correspond to the approximations, given by the right-hand sides of Eqs.\ (\ref{magic_equation_1}) and (\ref{magic_equation_2}). Additionally, in upper panels the thinnest lines correspond to $c_{\sigma}^2$. Upper panels: for both models, Eq.\ (\ref{magic_equation_1}) is valid for $T\lesssim100\ \txt{MeV}$ and $\mu_B\gtrsim600\ \txt{MeV}$. Lower panels: for both models, with the exception of the Walecka model at $T=200\ \txt{MeV}$, Eq.\ (\ref{magic_equation_2}) is valid for $\mu_B\gtrsim200\ \txt{MeV}$. Figure from \cite{Sorensen:2020ygf}.
	}
	\label{tests_of_formulas}
\end{figure}

We plot both sides of Eqs.\ (\ref{magic_equation_1}) and (\ref{magic_equation_2}) as functions of $\mu_B$ at a series of temperatures in Fig.\ \ref{tests_of_formulas}. More details on the expected behaviors of these quantities in both models can be found in \cite{Sorensen:2020ygf}. Here, comparing the exact results to the approximations, we see that Eq.\ (\ref{magic_equation_1}) is valid for $T\lesssim100\ \txt{MeV}$ and $\mu_B\gtrsim600\ \txt{MeV}$ (upper panels). On the other hand, the approximation introduced in Eq.\ (\ref{magic_equation_2}) is qualitatively valid for most of the probed $T$ and $\mu_B$, with the exception of regions characterized by $\mu_B\lesssim200\ \txt{MeV}$ (lower panels).

\section{Interpretation of experimental data and discussion}

\begin{figure}[t]
	\includegraphics[width = 0.99\columnwidth]{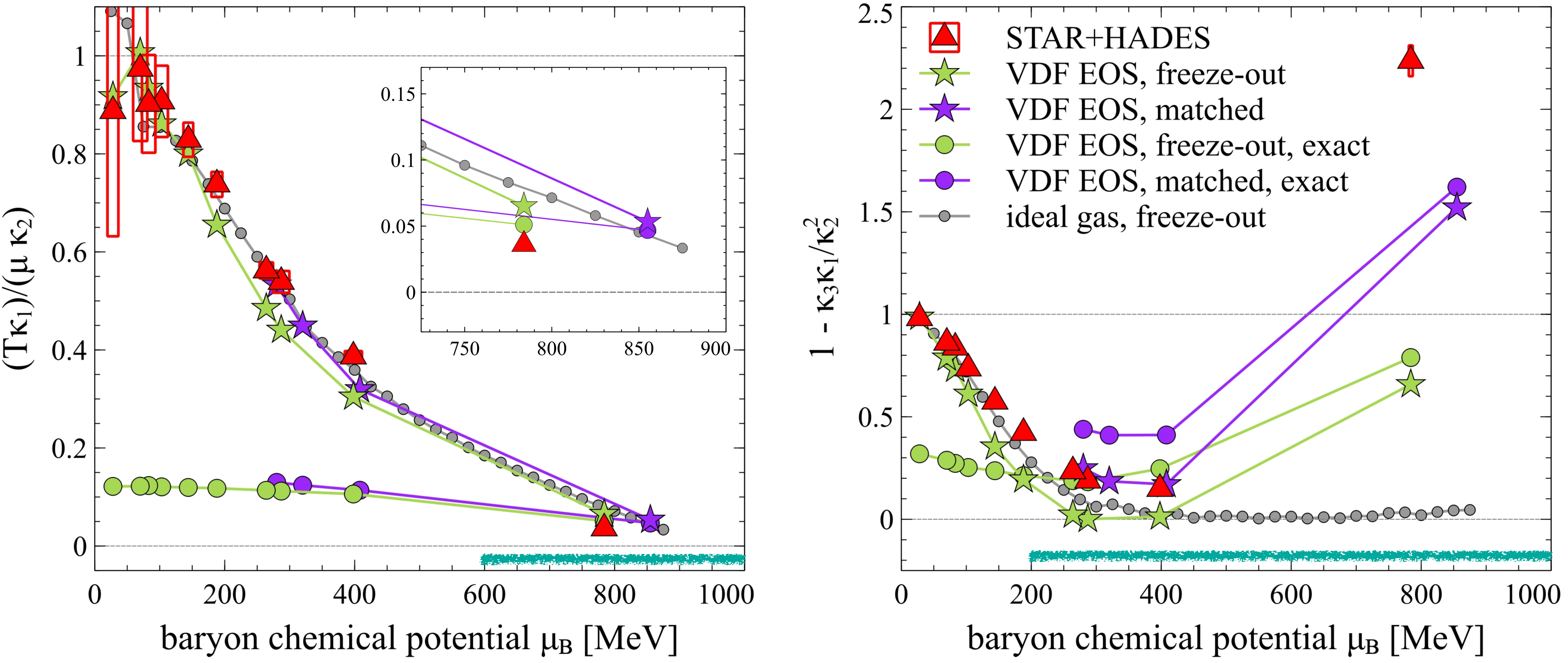}\\
	\caption{Comparison of the right-hand sides of Eq.\ (\ref{magic_equation_1}) (left panel) and Eq.\ (\ref{magic_equation_2}) (right panel) for experimental data (red triangles), ideal gas of nucleons along  the freeze-out parametrization line \cite{Cleymans:2005xv} (small gray circles), the VDF model at the freeze-out points (light green stars), and the VDF model at a set of points chosen to reproduce the data (dark purple stars); exact results, that is, the left-hand sides of Eqs.\ (\ref{magic_equation_1}) and (\ref{magic_equation_2}), are shown for the two cases considered in the VDF model (green and purple circles). The thick turquoise lines near the $\mu_B$-axis indicate the regions of validity of Eq.\ (\ref{magic_equation_1}) and Eq.\ (\ref{magic_equation_2}). The data points for the matched VDF results (shown only for collisions at low energies, where using the model is justified) are chosen to reproduce experimental values of $1-\kappa_3\kappa_1/\kappa_2^2$ (see Fig.\ \ref{diagram}). Figure adapted from \cite{Sorensen:2020ygf}.
	}
	\label{STAR_HADES_plots}
\end{figure}

We use experimental data on net proton cumulants and freeze-out parameters, as obtained by the STAR \cite{Abdallah:2021fzj} and HADES \cite{Adamczewski-Musch:2020slf, HADES_MLorentz_talk} experiments in collisions at 0-5\% centrality, to plot Eqs.\ (\ref{magic_equation_1}) and (\ref{magic_equation_2}) (red triangles, left and right panel in Fig.\ \ref{STAR_HADES_plots}, respectively) against $\mu_B$. The validity ranges of Eqs.\ (\ref{magic_equation_1}) and (\ref{magic_equation_2}), studied in the previous section, are indicated in Fig.\ \ref{STAR_HADES_plots} as thick turquoise lines near the $\mu_B$-axis. For the purposes of this initial exploration, we assume that the net proton number cumulant ratios can be used as a good proxy for the net baryon number cumulant ratios; we acknowledge this is likely inaccurate (see, e.g., \cite{Kitazawa:2011wh} for more discussion). We can see that in the case of Eq.\ (\ref{magic_equation_1}), only the experimental data for the lowest studied collision energy, at $\sqrt{s}=2.4\ \txt{GeV}$, can be trusted to indicate the value of $c_T^2$; as highlighted in the insert in the left panel of Fig.\ \ref{STAR_HADES_plots}, $c_T^2$ appears to be about half the value obtained for an ideal gas of nucleons at $T$ and $\mu_B$ calculated along the parametrization of the freeze-out line \cite{Cleymans:2005xv}. The results shown in the right panel of Fig.\ \ref{STAR_HADES_plots} indicate that the logarithmic derivative of $c_T^2$ decreases with decreasing collision energy, seemingly plateaus for $\sqrt{s_NN} = 7.7\ \txt{GeV}$, and then dramatically increases for $\sqrt{s}=2.4\ \txt{GeV}$. Together, these results indicate that in collisions at $\sqrt{s_{NN}} = 7.7\ \txt{GeV}$, $c_T^2$ is nearly constant as a function of $n_B$, while in collisions at $\sqrt{s}=2.4\ \txt{GeV}$, $c_T^2$ has a small value, but at the same time it is characterized by a large slope as a function of $n_B$. 

\begin{figure}[t]
	\begin{center}
		\includegraphics[width = 0.65\columnwidth]{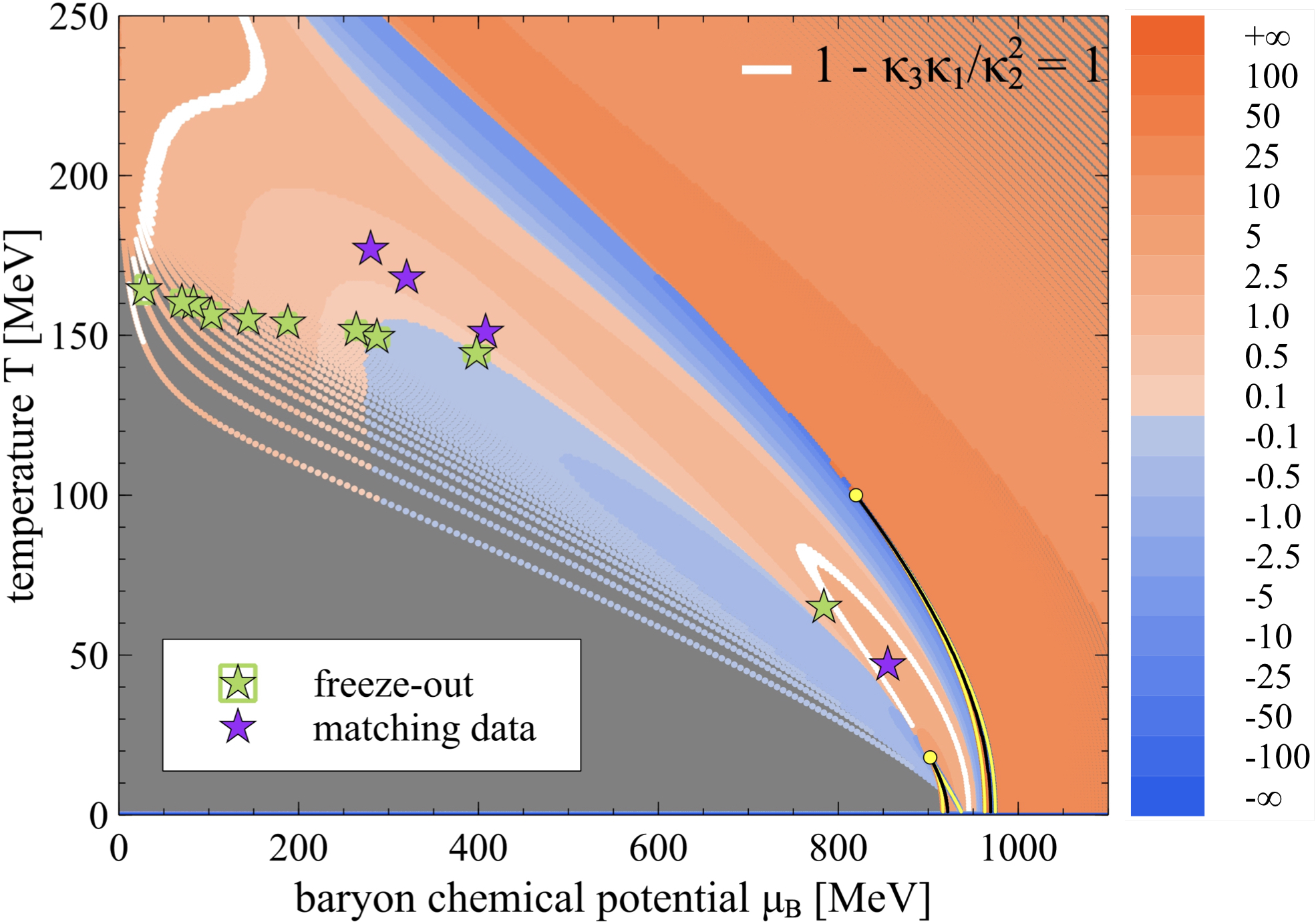}	
	\end{center}
	\caption{Contour plot of $1-\kappa_3\kappa_1/\kappa_2^2$ in the VDF model. Yellow and black lines correspond to the spinodal and coexistence lines, respectively; white contours signify regions where $1-\kappa_3\kappa_1/\kappa_2^2=1\pm0.03$. Light green stars denote experimentally measured freeze-out parameters $(T_{\txt{fo}},\mu_{\txt{fo}})$, while dark purple stars denote points where $1-\kappa_3\kappa_1/\kappa_2^2$, taken along lines informed by average phase diagram trajectories for STAR collision energies \cite{Shen:2020jwv}, matches the experimentally measured values for a given collision energy. The softening of the EOS, leading to negative values of $\big(\infrac{d\ln c_T^2}{d\ln n_B}\big)_T$, occurs in two regions of the phase diagram, corresponding to the ordinary nuclear matter phase transition and to the conjectured QGP-like phase transition. Figure from \cite{Sorensen:2020ygf}.
	}
	\label{diagram}
\end{figure}

To get a better intuition about the thermodynamics that might be at play here, we study the $T$- and $\mu_B$-dependence of $1 - \kappa_3 \kappa_1/\kappa_2^2$ in the VDF model, shown in a diagram form in Fig.\ \ref{diagram}. We find points, denoted in the Figure by purple stars, at which the values of $1 - \kappa_3 \kappa_1/\kappa_2^2$ match the experimentally measured values for a given collision energy; for collisions at STAR energies, this search is additionally guided by the average phase diagram trajectories of the collision systems \cite{Shen:2020jwv}. By comparing with Fig.\ \ref{tests_of_formulas}, we see that at the point reproducing the experimental value of $1-\kappa_3\kappa_1/\kappa_2^2$ for the lowest STAR energy, $c_T^2$ is nearly constant as a function of $\mu_B$ (thick short-dashed line, $T=150\ \txt{MeV}$, at $\mu_B\approx410\ \txt{MeV}$ in Fig.\ \ref{tests_of_formulas}), while at the point reproducing the result for the HADES energy, $c_T^2$ increases sharply with $\mu_B$ (thick solid line, $T=50\ \txt{MeV}$, at $\mu_B\approx850\ \txt{MeV}$ in Fig.\ \ref{tests_of_formulas}), agreeing with conclusions based on the behavior of $1 - \kappa_3 \kappa_1/\kappa_2^2$.

The values of both $(T\kappa_1)/(\mu_B \kappa_2)$ and $1 - \kappa_3 \kappa_1/\kappa_2^2$ obtained in the VDF model at the ``matched'' $(T,\mu_B)$ points are plotted in Fig.\ \ref{STAR_HADES_plots} (purple stars), where we also show results corresponding to values of $T$ and $\mu_B$ equal to the freeze-out parameters (green stars); additionally, we show the exact values of $c_T^2$ and $c_T^2 + \big(\infrac{d\ln c_T^2}{d\ln n_B}\big)_T$ as obtained in the VDF model at these points (purple and green circles). Here, we can easily see that values of $1 - \kappa_3 \kappa_1/\kappa_2^2$ obtained at $(T,\mu_B)$ given by the freeze-out parameters (green stars) do not lead to a good agreement with data. At the same time, the ``matched'' points (purple stars) can be clearly interpreted as corresponding to regions of the phase diagram probed before the freeze-out (see Fig.\ \ref{diagram}). Altogether, this suggest that the measured values of the cumulants are affected by stages of the collision preceding the freeze-out; indeed, it is known that critical fluctuations exhibit a large relaxation time \cite{Berdnikov:1999ph,Stephanov:2017ghc,Du:2020bxp}, which may support this interpretation of the results of our model comparison.

We want to stress here that whether the values of the cumulants are indeed not equilibrated at freeze-out and therefore reflect the state of the system at earlier stages of the evolution should be further investigated (recent studies pointing to such possibility can be found in \cite{Nahrgang:2018afz,Jiang:2017sni}). Moreover, experiments measure not baryon number cumulants, as used in our model, but proton number cumulants. Effects due to baryon number conservation may also be important \cite{Bzdak:2012an,Vovchenko:2020gne}.

Nevertheless, our initial study within the VDF model, which is well-justified for low-energy collisions whose evolution is dominated by the hadronic stage, suggests that collisions at the lowest STAR and HADES energies may be probing regions of the phase diagram where the cumulants tell us more about hadronic physics than the QCD CP. In particular, the change in the sign of $\kappa_3$, predicted to take place in the vicinity of a critical point \cite{Asakawa:2009aj} and apparent in the HADES data (see right panel of Fig.\ \ref{STAR_HADES_plots}), may mark the region of the phase diagram affected by the nuclear liquid-gas phase transition. If this is true, it may be worthwhile to study the cumulants at even lower collision energies, starting from $0.1\ \txt{GeV}$ projectile kinetic energy, and attempt to obtain $c_T^2$ around the nuclear liquid-gas CP. Conversely, at higher energies it could be possible to use collisions at different centralities and rapidity windows to estimate the neglected terms in Eqs.\ (\ref{magic_equation_1}) and (\ref{magic_equation_2}), and obtain a stronger estimate for $c_T^2$ in the respective regions of the phase diagram. Further studies of effects due to dynamics, in particular using simulations, will be absolutely essential in determining the extent to which the proposed method provides a reliable extraction of $c_T^2$ and its derivatives.

\vspace{10mm}

\noindent\textbf{\large{Acknowledgments}} \vspace{3mm}

A.S.\ and V.K.\ received support through the U.S. Department of Energy, 
Office of Science, Office of Nuclear Physics, under contract no.\ 
DE-AC02-05CH11231231 and received support within the framework of the
Beam Energy Scan Theory (BEST) Topical Collaboration. D.O.\ and L.M.\ were supported by the U.S.\ DOE under Grant No.\ DE-FG02-00ER4113.

\end{document}